\documentclass[preprint,showpacs,amsmath,amssymb,byrevtex,endfloats*,jasa,altaffilletter]{revtex4asa}

\usepackage[pdftex]{graphicx}
\usepackage{epstopdf}
\usepackage{dcolumn}
\usepackage{bm}
\usepackage{helvet} 
\usepackage{epsfig} 

\begin{document}


\title[
Porosity dependence of sound speed
]{%
Sound speed in water-saturated glass beads as a function of frequency and porosity}


\author{Theodore F. Argo IV%
}
\defineheaderauthor{Argo IV}
\author{Matthew D. Guild%
}
\author{Preston S. Wilson%
}
\affiliation{Department of Mechanical Engineering and Applied Research Laboratories; The University of Texas at Austin\\
 P.O. Box 8029; Austin TX 78713-8029%
}

\author{Matthias Schr\"{o}ter%
}
\affiliation{Max Planck Institute for Dynamics and Self-Organization\\
 Bunsenstra\ss e 10; 37073 G\"{o}ttingen, Germany%
}

\author{Charles Radin%
}
\email{radin@math.utexas.edu}
\affiliation{Department of Mathematics, The University of Texas at Austin, 1 University Station C1200; Austin TX 78712%
}

\author{Harry L. Swinney%
}
\affiliation{Center for Nonlinear Dynamics and Department of Physics, The University of Texas at Austin, 1 University Station C1610; Austin TX 78712%
}

\defineshorttitle{Porosity dependence of sound speed}
\definerunningtitle{Porosity dependence of sound speed}

\date{\today}

\begin{abstract}
Sound propagation in water-saturated granular sediments is known to depend on the sediment porosity, but few data in the literature address both the frequency and porosity dependency.  To begin to address this deficiency, a fluidized bed technique was used to control the porosity of an artificial sediment composed of glass spheres of 265~$\mu$m-diameter. Time-of-flight measurements and the Fourier phase technique were utilized to determine the sound speed for frequencies from 300~kHz to 800~kHz, and porosities from 0.37 to 0.43. A Biot-based model qualitatively describes the porosity dependence.
\end{abstract}

\pacs{%
43.30.Ma, 43.35.Bf}

\maketitle

\section{Introduction}
Numerous models have been proposed to describe sound propagation in water saturated granular materials (see Refs.~[\onlinecite{Biot:1962,Schwartz:1984,Stoll:1989,Buckingham:1998,Chotiros:2004}] for a sampling); however, no model yet accurately describes the sound speed in these materials across the full range of frequencies of interest in underwater acoustics.  Measurements of the acoustic properties of water saturated sediments have been made under a variety of conditions including \textit{in situ} measurements,\cite{Turgut:1990,Williams:2002} retrieved samples,\cite{Hamilton:1982,Williams:2002} reconstituted samples,\cite{Shirley:1978,Lee:2007} and artificial sediments.\cite{Hefner:2006} There is a large variability in the observed sound speeds of these various sediments due to the wide range of porosities, permeabilities, and other physical properties they exhibit.  Richardson and Jackson have studied the effect of porosity on a global scale, compiling numerous measurements obtained across many sediment types.\cite{Jackson:2007}  Naturally, the sediment porosity changes are also accompanied by grain size and material differences.

In none of these studies, nor in any other studies known to the authors, was the effect of porosity studied in a controlled manner across a range of frequencies, for a single sediment sample with uniform grain size distribution.  In such an idealized case, one can focus on the effect of porosity in isolation, and, with sufficient knowledge of most of the sediment physical parameters, one can make meaningful comparisons to various models of sound propagation.  Towards this end, a fluidized bed apparatus was used to prepare specific porosities of a spherical glass bead sediment, and ultrasonic time-of-flight measurements were obtained.  Experimental results were compared to the predictions of the Williams effective density fluid model (EDFM).\cite{Williams:2001}  The purpose of this brief communication is to describe the porosity-control technique as applied to sediment acoustics and to present a set of example measurements.

\section{Description of Experiment}

\begin{figure}
\ifgalleyfig
  \includegraphics[width=1\columnwidth]{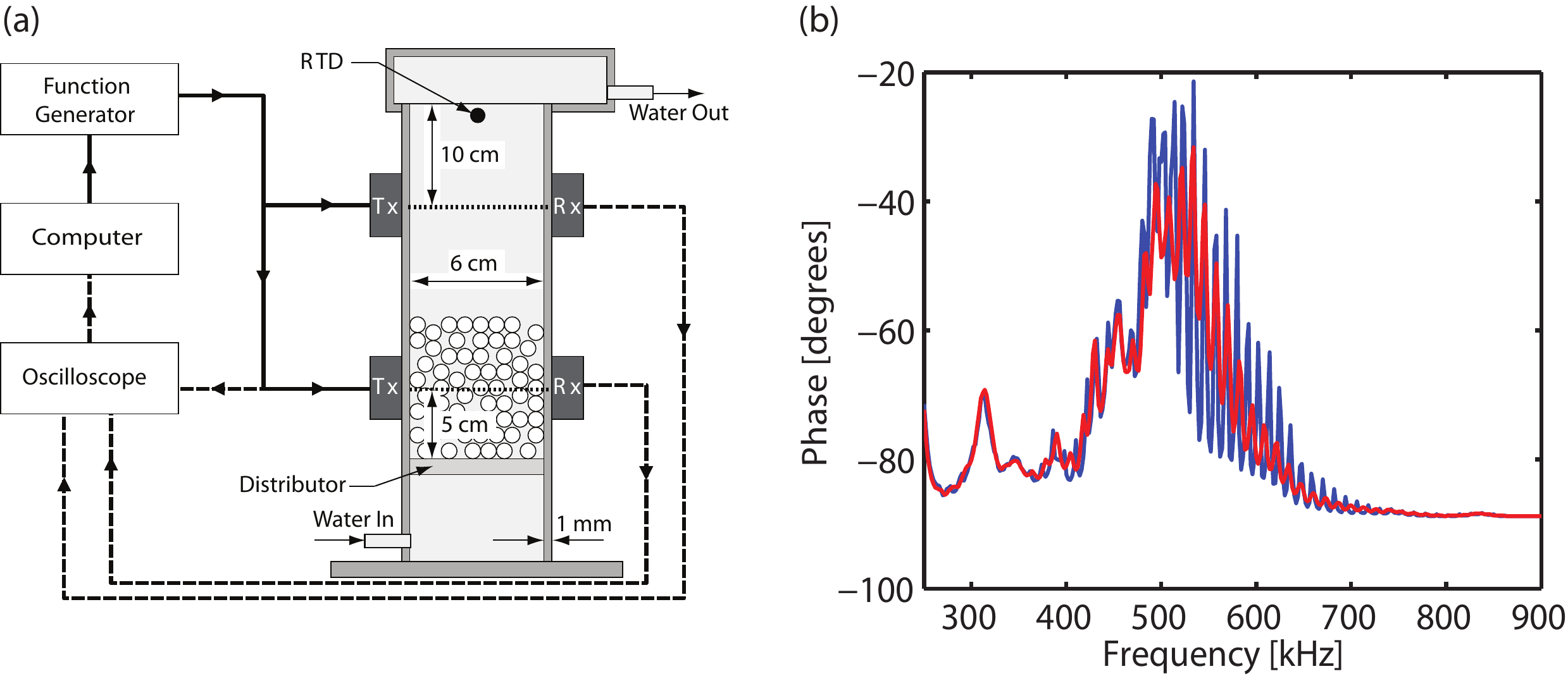}
\else
  \includegraphics[width=1\columnwidth]{Figure1V4.eps}
\fi
\caption[(Color online) A schematic diagram of the experimental apparatus is shown in (a).  Solid (dashed) lines denote the transmitted (received) electrical signal paths.  The glass beads are not drawn to scale. The phase of the measured electrical input impedance of one of the receiving transducers depicted in (a) is shown in (b).  The blue curve is for the transducer mounted in the apparatus when filled with water, and the red curve is for water-saturated beads.  The phase difference apparent within the experimental frequency range (300~kHz through 800~kHz) was found to lead to correctable systematic errors in the sound speed measurement.]
{\label{fig1}
\ifgalleyfig{(Color online) A schematic diagram of the experimental apparatus is shown in (a).  Solid (dashed) lines denote the transmitted (received) electrical signal paths.  The glass beads are not drawn to scale. The phase of the measured electrical input impedance of one of the receiving transducers depicted in (a) is shown in (b).  The blue curve is for the transducer mounted in the apparatus when filled with water, and the red curve is for water-saturated beads.  The phase difference apparent within the experimental frequency range (300~kHz through 800~kHz) was found to lead to correctable systematic errors in the sound speed measurement.}\fi}
\end{figure}

The artificial sediment was composed of soda lime glass spheres of $265$~$\mu$m $\pm 15$~$\mu$m diameter saturated with degassed distilled water.  The beads were sorted using a sieve shaker between 250~$\mu$m and 280~$\mu$m fine mesh sieves.  The mass of the dry spheres was measured, and the spheres were then added to the apparatus shown in Fig.~\ref{fig1}a. Degassed distilled water was then continuously pumped through the sample for $30$ minutes and then allowed to rest overnight to dissolve trapped gas bubbles.

Sediment porosity $\beta$ was controlled by using the fluidized bed apparatus shown in Fig.~\ref{fig1}a, which is similar to that described by Schr\"{o}ter \textit{et al.}\cite{Schroeter:2007} The interior of the sample holder was $6$~cm $\times$ $6$~cm square and $30$~cm tall, with the sediment height dependent upon porosity, but nominally $100$~mm.  Two parallel polycarbonate walls of the sample holder were $1$~mm thick and the opposite two were $10$~mm thick.  Control of the porosity of the monodisperse fluid-saturated bead pack is initiated by pumping water at a particular flow rate through the distributor (a block of sintered bronze powder, 5~mm thick with approximately 10~$\mu$m-diameter pores).  The beads are carried into suspension, with the overall magnitude of their motion governed by the flow rate.  The beads achieve steady-state motion and after two minutes, the flow is stopped.  The beads are then given one minute to settle.  The resulting porosity is dependent upon the flow rate.  An additional procedure was used to achieve uniform porosity for $\beta < 0.40$:  A series of flow pulses were introduced, starting at a high flow rate but with an incrementally decreasing flow rate.  This procedure allowed the bead column to settle more uniformly.  The spatial variability of the porosity achieved by this technique was measured using micro-computed x-ray tomograms and porosity variations of about $\pm0.005$ were observed.

A rectangular excitation pulse with a 10~volt amplitude and a 1.25~$\mu$s width was directed to a pair of broadband ultrasonic transducers (labeled~Tx) with center frequencies of 500~kHz, which transmitted simultaneously through the water and sediment paths, as shown in Fig.~\ref{fig1}a.  Another pair of ultrasonic transducers (labeled~Rx) with center frequencies of $550$~kHz received the acoustic signals. Commercially-available ultrasonic coupling gel was used between the transducer faces and the sample holder walls and custom clamps were used to repeatably locate and hold the transducers in place.  Their positions were held constant during the course of the experiments.  A digital storage oscilloscope acquired both the source signal and the received signals.  In all cases, $512$ linear time-domain signals were averaged onboard the scope to remove noise.  The averaged signals were transferred to a computer for analysis.

\label{porsec}

To determine sediment porosity, images of the beads were obtained with a digital camera at the same time as the acoustic measurements.  A fiducial scale indicating sediment height $h$ was placed within the field of view of the camera.  The porosity of the sample $\beta$ was determined with $\beta=1-m_{\mathrm{b}}/\left(\rho_{\mathrm{b}}hA\right)$, where $m_{\mathrm{b}}$ and $\rho_{\mathrm{b}}$ are the dry mass and density of the beads, $h$ is the height of the beads within the column, and $A$ is the cross-sectional area of the column.  The uncertainty in the porosity is primarily due to the uncertainty in the area and the bead density.\cite{Schroeter:2005}  In the present experiment, this resulted in a porosity uncertainty of $\pm$0.003.   The packing state of the sediment was found to be unaffected by the acoustic excitation.  This was verified by comparing pings at the beginning, middle, and end of the acoustic averaging interval.  No systematic variations in the received waveforms were found.
A resistance temperature detector, labeled RTD in Fig.~\ref{fig1}a, was used to measure the temperature during the acoustic measurements. Equation $5.6.8$ of Ref.~[\onlinecite{Kinsler:1999}] was later used to determine the sound speed $c_{0}$ of the distilled water at the experimental temperature.

The apparatus and preparation technique described above yielded porosity control but also resulted in an acoustic path length that was shorter than desired.  For a fixed path length uncertainty, sound speed measurement uncertainty is inversely proportional to path length.  Increasing the cross-sectional area of the apparatus while retaining uniform porosity is difficult. Since the goal of this work was to investigate the effect of porosity, we chose to achieve a homogeneous sediment at the expense of simplicity in the acoustic data analysis.  In addition, the porosity control procedure can take as long as several hours for the dense sediments.  Because the water temperature follows temperature fluctuations in the laboratory, there could be significant temperature-dependent sound speed changes in the sample before and after the procedure.  Hence, the four-transducer apparatus of Fig.~\ref{fig1}a was used to facilitate contemporaneous post-preparation water-path calibration and sediment-path measurements.

Coupling between the sediment sample and the tank walls and transducers was found to be significant, as demonstrated in Fig.~1b, which shows the phase of the measured electrical input impedance of one of the receiving acoustic transducers when mounted in the apparatus filled only with water, and when filled with water-saturated beads.  There is as much as a thirty degree phase difference between these two cases.  Further, initial measurements with distilled water revealed that the sound speed determined by Eq.~\ref{TOFeq} applied to the present four-transducer apparatus included a systematic error due to a small path length difference between the two transducer pairs.  The effect of these systematic errors was found to be the same order of magnitude (a few percent) as the sound speed changes expected from the porosity dependency. Therefore, the calibration and error correction procedure described in Section~\ref{sysmod} was implemented to overcome these path length and coupling-related errors. Near-field effects were quantified as described in Ref.~[\onlinecite{Xu:1993}] and found to be insignificant, which was also the case for a similar experiment described in Ref.~[\onlinecite{Lee:2007}].

\section{Analysis}

The Fast Fourier Transforms (FFTs) of the received signals for both the water-paths and sediment-paths were calculated.  A Fourier phase technique\cite{Molyneux:1999} was then used to obtain a measurement of the apparent sound speed $c_{\mathrm{sys}}$ of the sediment within the experimental system.  This sound speed was calculated with
\begin{equation}
\label{TOFeq}
c_{\mathrm{sys}} = c_{0} \left[1-\frac{c_{0} \Delta \phi}{l\omega}\right]^{-1},
\end{equation}
where $c_{0}$ is the sound speed in the water portion of the fluidized bed determined by Eq.~5.6.8 of Ref.~[\onlinecite{Kinsler:1999}], $l$ is the length of travel for the acoustic wave within the sample, $\omega =  2\pi f$ is the angular frequency, and $\Delta\phi$ is the difference in phase between the two FFT spectra at frequency~$\omega$.

\label{sysmod}

\begin{figure}
\ifgalleyfig
  \includegraphics[width=1\columnwidth]{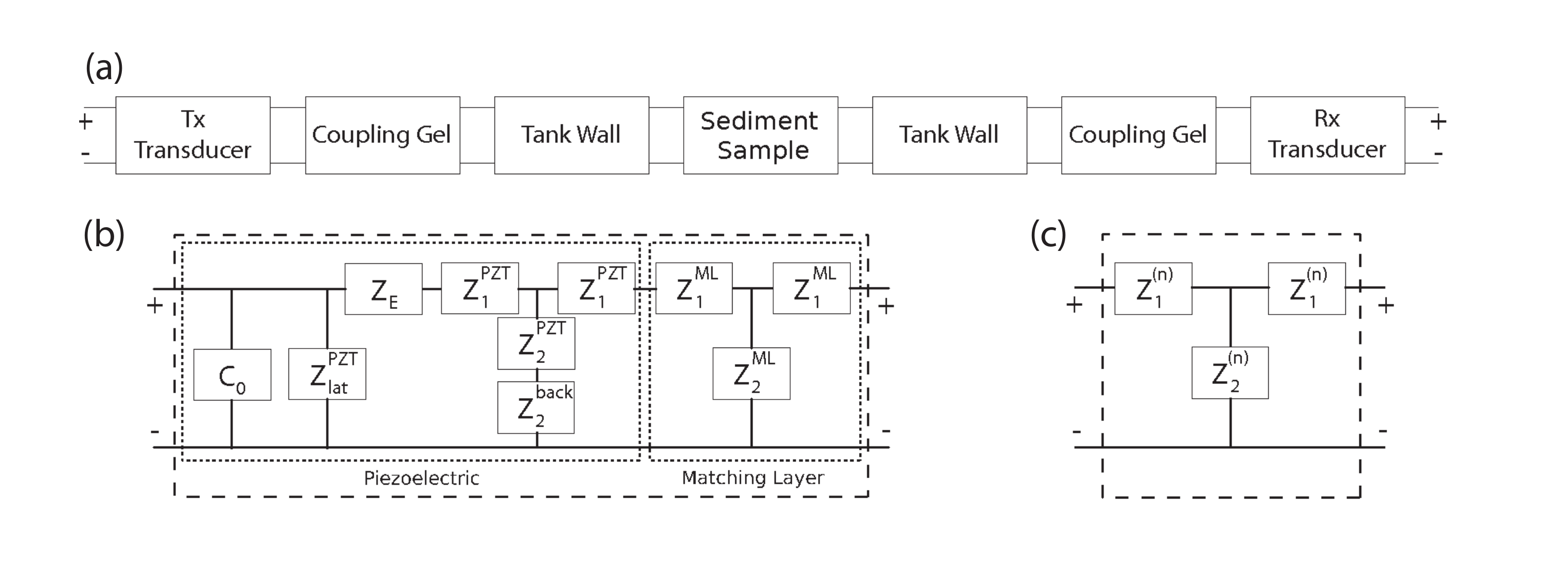}
\else
  \includegraphics[width=1\columnwidth]{X20090511_xducer_model_design2}
\fi
\caption[A schematic of the transmission line model is shown in (a).  It is composed of a Mason model for the transducers~(b) and generic transmission line elements (c) for the remaining layers.]
{\label{fig2}
\ifgalleyfig{A schematic of the transmission line model is shown in (a).  It is composed of a Mason model for the transducers~(b) and generic transmission line elements (c) for the remaining layers.}\fi}
\end{figure}

An equivalent circuit transmission line model of the experimental system (Fig.~\ref{fig2}a) was used to correct for the systematic errors described in Sec.~\ref{porsec}.   The model included the transducers, coupling gel, tank walls, and the sediment (or water).  The passive layers (Fig.~\ref{fig2}c) were modeled with $Z_{0}^{\left(n\right)}=\rho^{\left(n\right)} c^{\left(n\right)} S^{\left(n\right)}/N^{2}$, $Z_{1}^{\left(n\right)}=jZ_{0}^{\left(n\right)}\tan{\left[k^{\left(n\right)}l^{\left(n\right)}/2\right]}$, and $Z_{2}^{\left(n\right)}=-jZ_{0}^{\left(n\right)}/\sin{\left[k^{\left(n\right)}l^{\left(n\right)}\right]}$, where $k^{\left(n\right)}$, $\rho^{\left(n\right)}$, and $c^{\left(n\right)}$ are the wave number, density, and sound speed within the $n$th component respectively, $l^{\left(n\right)}$ is the path length in the $n$th component,  $S^{\left(n\right)}$ is the surface area, and $N$ is the transformation factor from one physical domain to another.  The frequency-dependent value of the sediment sound speed used in the system model is denoted by $c_{\mathrm{sed}}$ and is used to construct both $k$ and $c$ for the sediment layer.  A Mason equivalent circuit model of a thickness mode piezoelectric resonator with a matching layer \cite{Mason:1958} was used to model the transducers (Fig.~\ref{fig2}b).  The unloaded electrical impedance of each transducer was measured and the values for each model element shown in Fig.~\ref{fig2}b were found by fitting the model to the impedance measurement, including the use of the properties for the piezoelectric ceramic PZT-4 listed in Table $4.3$ of Ref.~[\onlinecite{Wilson:1988}].  Additional material properties for the model elements are listed in Table~\ref{mat_props}.  Measurements made with distilled degassed water occupying both transmission paths allowed for calibration of the transmission path lengths in the model.

To perform the correction, the experimental input signals for the water and sediment paths were digitized, FFTs were taken, the resulting spectra were input into the system model, and the output spectra were calculated.  These modeled output spectra were then subjected to the calculation described in Eq.~\ref{TOFeq} which yield a modeled system value of  $c_{\mathrm{sys,m}}$.  Using Newton-Raphson iteration, $c_{\mathrm{sed}}$ was systematically varied in the model until the modeled system value $c_{\mathrm{sys,m}}$ matched the measured value of $c_{\mathrm{sys}}$ for each frequency $\omega$ and for each porosity $\beta$.  The values of $c_{\mathrm{sed}}$ that achieved this match were taken to be the free-field sound speed of the sediment samples and are reported in Figs.~\ref{fig3}b and \ref{fig4}.

Imperfect knowledge of the water and sediment path lengths, known to within 0.1~mm, and total length of the other components (pzt, matching layer, coupling gel and sample holder wall), known to within 0.05~mm, were the largest source of measurement uncertainty in this work.  The effect of these uncertainties was determined using the circuit model and resulted in a $\pm$0.0015 uncertainty in sound speed ratio, which is about the same size as the data points used in Figs.~3b and 4.

\section{Results}

\begin{figure}
\ifgalleyfig
 		\includegraphics[width=1\columnwidth]{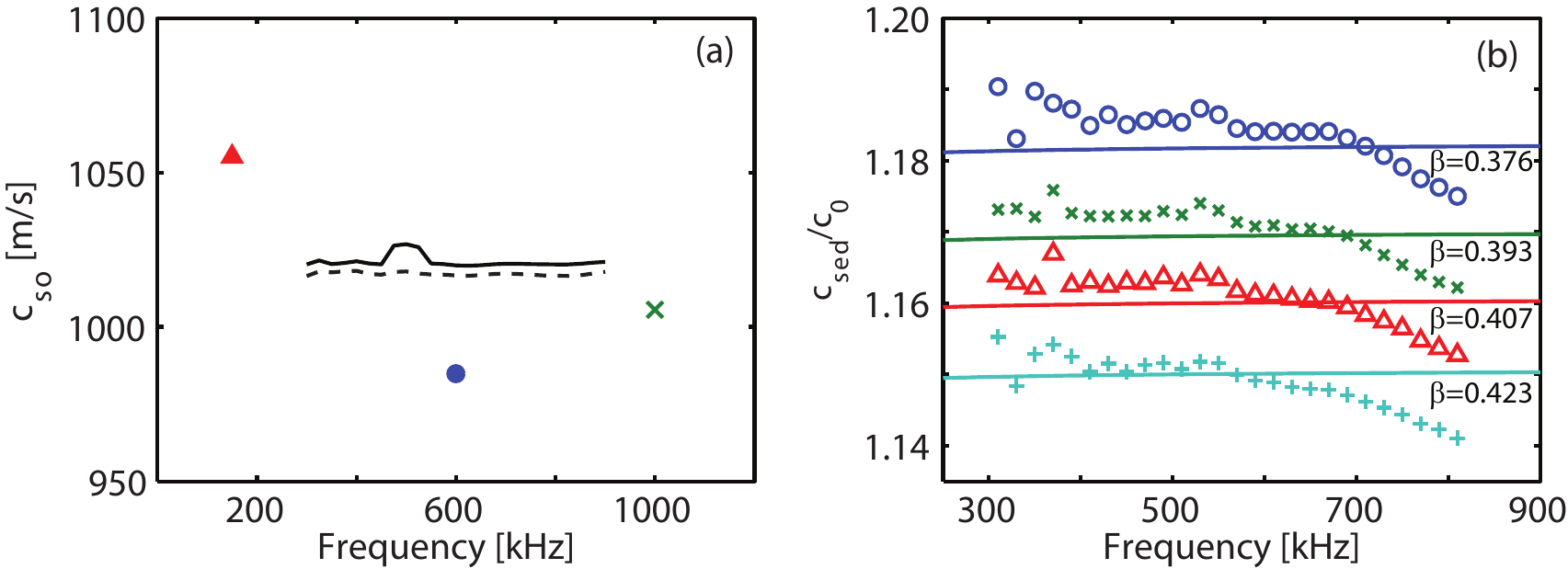}
\else
  		\includegraphics[width=1\columnwidth]{Figure3V2}
\fi
\caption[(Color online) The measured sound speed for silicone oil samples are shown as a function of frequency in (a).  The solid line corresponds to $1\times10^{5}$~cSt silicone oil at 23.0$^{\circ}$C and the dashed line corresponds to $1\times10^{6}$~cSt silicone oil at 23.6$^{\circ}$C.  The $\times$ corresponds to the measurement of sound speed in 100~cSt silicone oil from Ref.~\onlinecite{Hartmann:1974}, the circle corresponds to the measurement of sound speed in 1000~cSt silicone oil from
Ref.~\onlinecite{Weissler:1949}, and the triangle corresponds to the measurement of sound speed in 1000~cSt silicone oil from Ref.~\onlinecite{Capps:1981}.  The literature sound speeds were extrapolated to $23.0^{\circ}$C using the temperature coefficients stated in their respective references. Measured sound speed ratios of water-saturated glass beads are shown as a function of frequency in (b).  Solid lines correspond to the predictions of the EDFM\cite{Williams:2001} and symbols correspond to measurements.  In (b), circles correspond to a porosity of $0.376$, $\times$s correspond to a porosity of $0.393$, triangles correspond to a porosity of $0.407$, and crosses correspond to a porosity of $0.423$.]
{\label{fig3}
\ifgalleyfig{(Color online) The measured sound speed for silicone oil samples are shown as a function of frequency in (a).  The solid line corresponds to $1\times10^{5}$~cSt silicone oil at 23.0$^{\circ}$C and the dashed line corresponds to $1\times10^{6}$~cSt silicone oil at 23.6$^{\circ}$C.  The $\times$ corresponds to the measurement of sound speed in 100~cSt silicone oil from Ref.~\onlinecite{Hartmann:1974}, the circle corresponds to the measurement of sound speed in 1000~cSt silicone oil from
Ref.~\onlinecite{Weissler:1949}, and the triangle corresponds to the measurement of sound speed in 1000~cSt silicone oil from Ref.~\onlinecite{Capps:1981}.  The literature sound speeds were extrapolated to $23.0^{\circ}$C using the temperature coefficients stated in their respective references. Measured sound speed ratios of water-saturated glass beads are shown as a function of frequency in (b).  Solid lines correspond to the predictions of the EDFM\cite{Williams:2001} and symbols correspond to measurements.  In (b), circles correspond to a porosity of $0.376$, $\times$s correspond to a porosity of $0.393$, triangles correspond to a porosity of $0.407$, and crosses correspond to a porosity of $0.423$.}\fi}
\end{figure}

\begin{figure}
\ifgalleyfig
  \includegraphics[width=1\columnwidth]{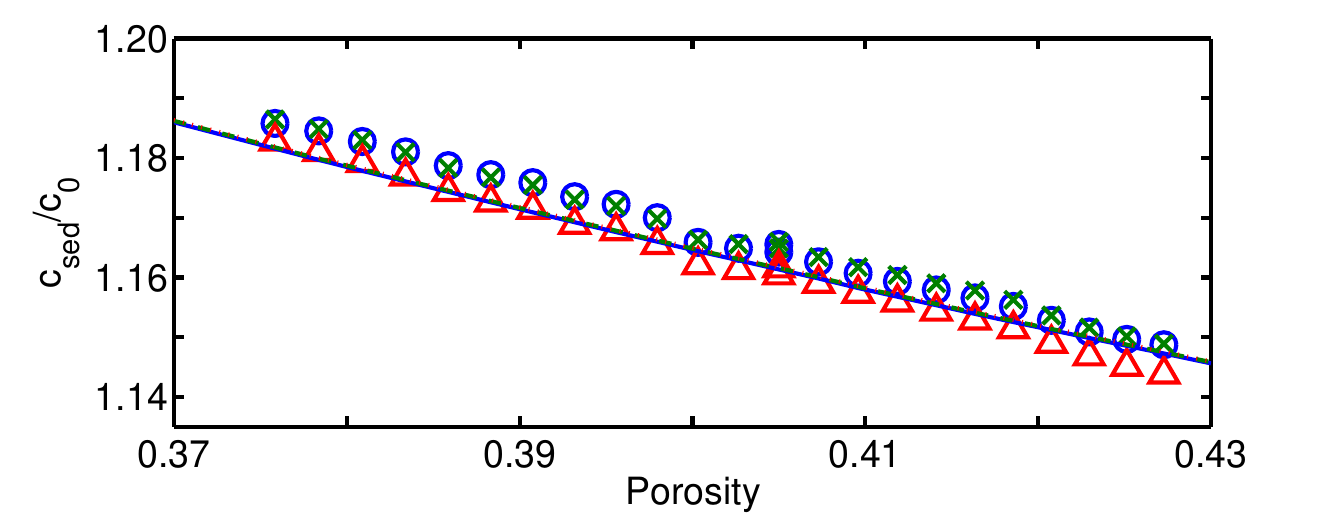}
\else
  \includegraphics[width=1\columnwidth]{Figure4V5.eps}
\fi
\caption[Measured sound speed ratios of water-saturated glass beads are shown as a function of porosity. Dashed and solid lines (nearly indistinguishable) correspond to the predictions of the EDFM\cite{Williams:2001} and symbols correspond to measurements.  Circles correspond to a frequency of $400$~kHz, $\times$s correspond to a frequency of $550$~kHz, and triangles correspond to a frequency of $700$~kHz.]
{\label{fig4}
\ifgalleyfig{Measured sound speed ratios of water-saturated glass beads are shown as a function of porosity. Dashed and solid lines (nearly indistinguishable) correspond to the predictions of the EDFM\cite{Williams:2001} and symbols correspond to measurements.  Circles correspond to a frequency of $400$~kHz, $\times$s correspond to a frequency of $550$~kHz, and triangles correspond to a frequency of $700$~kHz.}\fi}
\end{figure}

A calibration of the system was performed using silicone oil with viscosities of $1\times10^{5}$ and $1\times10^{6}$ cSt. The calibration was performed by adding a 10~cm thick layer of silicone fluid to the top of a 10~cm thick layer of degassed distilled water.  Since silicone fluid is less dense than water and the two fluids are immiscible, stratification of the fluids was achieved.  Sound speeds were calculated using Eq.~\ref{TOFeq} and processed using the model outlined in Section \ref{sysmod}.  The resultant sound speeds, $c_{\textrm{so}}$, are shown in Fig.~\ref{fig3}a and fall within a range of values found in the literature.\cite{Hartmann:1974,Weissler:1949,Capps:1981}

In Fig.~\ref{fig3}b, the frequency dependence of the corrected sound speed is compared to the predictions of the Effective Density Fluid Model (EDFM),\cite{Williams:2001} for the parameters given in Table \ref{EDFMparam}.  Negative dispersion is apparent in Fig.~\ref{fig3}b for all porosities above 550~kHz.  Considering just one frequency, say 600~kHz, the variation in the measured sound speed as a function of porosity is well-predicted by the EDFM.

In Fig.~\ref{fig4}, the porosity dependence of the measured sound speed is shown and compared to the predictions of the EDFM at three frequencies.  Measurements at porosities not shown in Fig.~\ref{fig3}b are shown here.  It is apparent that the EDFM does a good job of describing the observed relative porosity dependence.  Note that all of the sound speeds predicted by the EDFM nearly collapse onto a single line.  The negative dispersion exhibited by the measured sound speed remains apparent.  Plotted this way, the sound speed monotonically decreases with porosity. Therefore, the porosity appears to be a good indicator of the sound speed in this glass bead sediment.

\section{Conclusion}

The porosity- and frequency-dependent sound speed in an artificial water-saturated glass bead sediment was measured using a fluidized bed apparatus with a time-of-flight acoustic measurement and a Fourier phase technique.  The relative dependence of the sound speed on porosity is properly predicted by the Biot-based EDFM model throughout the experimental frequency range, 300~kHz to 800~kHz, and it also quantitatively described the sound speed ratio to within about $\pm 1\%$.   The measured frequency dependence reveals negative dispersion at frequencies above $550$~kHz.   The authors acknowledge that other models that address multiple scattering~\cite{Schwartz:1984} and high-frequency viscosity effects~\cite{Chotiros:2008} will likely do a better job predicting the dispersion in this frequency range.  The present purpose is to report the porosity control technique and initial measurements obtained with it, not to identify the physical cause of the observed dispersion.  The reported measurement and data analysis techniques are similar to those reported in Refs.~\onlinecite{Schwartz:1984} and \onlinecite{Lee:2007} in which negative dispersion was also observed.  All of these techniques are variations of the common immersion technique and there is no evidence indicating that the determination of sound speed with these techniques is dependent on the physical nature or origin of the dispersion.  Measurements with a variety of bead sizes and with natural sand grains, and a more thorough model evaluation are currently underway.

\begin{table}[fp]
\begin{tabular}{c|c|c}
\hline
 				&	density (kg/m$^3$)	&	sound speed (m/s)			\\\hline
matching layer		&	1200				&	$2650\left(1+0.01j\right)$		\\\hline
coupling gel		&	1000				&	$1500$					\\\hline
sample holder wall	&	1200				&	$1800\left(1+0.01j\right)$		\\\hline
\end{tabular}
\caption{Parameters used in the transmission line model. The sound speed and density of the transducer matching layers and coupling gel were obtained from their respective manufacturers.  The sample holder wall density was measured, and its sound speed was also measured using commercial ultrasonic testing instrumentation.\label{mat_props}
}
\end{table}

\begin{table}[fp]
\begin{tabular}{c|c|c|c}
\hline
grain density 				&	2487 $\pm 1$	 kg/m$^{3}$		& water density 			&	998 kg/m$^{3}$						\\\hline
bead bulk mod.				&	41.175 GPa			& water sound vel. (m/s)		&	1474, 1472, 1471, 1472 					\\\hline
water viscosity				&	0.001 Pa$\cdot$s		& permeability 	(m$^{-2}$)	&	[5.32, 6.44, 7.50, 8.87]$\times10^{-11}$	\\\hline
tortuosity 					&	1.63, 1.59, 1.57, 1.54	& porosity					&	0.376, 0.393, 0.407, 0.423			\\\hline
\end{tabular}
\caption{Parameters used as inputs to the Williams EDFM model.\cite{Williams:2001}  Grain density was measured using a Gay-Lussac flask.  Water sound speed was given by Eq.~$5.6.8$ in Ref.~[\onlinecite{Kinsler:1999}]. Tortuosity was calculated using Ref.~[\onlinecite{Dias:2006}] with an exponent of $0.5$.  Permeability was calculated using the Kozeny-Carman Equation\cite{Batu:1998} with a coefficient of $5$ and the pore size factor given by Ref.~[\onlinecite{Williams:2001}].  Bead bulk modulus was taken from the manufacturer's specification. Porosity was determined by the method in Sec.~\ref{porsec}. When multiple entries appear within a cell, they are for the four porosities that are tabulated here, respectively, and that also appear in Fig.~\ref{fig3}b.\label{EDFMparam}}
\end{table}

\begin{acknowledgments}
This work was supported by the Office of Naval Research Ocean Acoustics Office, the Robert A.\ Welch Foundation Grant F-0805, and National Science Foundation Grant DMS-0700120.  The authors also wish to acknowledge Nicholas P. Chotiros for insightful discussions and Udo Krafft for experimental support.
\end{acknowledgments}

\bibliographystyle{jasanum}
\bibliography{JASAEL_Argo_Guild_draft_v4b}

\maketablesandfigures
\end{document}